\documentclass[aps,prl,showpacs,twocolumn,superscriptaddress]{revtex4}
\usepackage{color}
\usepackage{bm}
\usepackage{amsmath}
\usepackage{amssymb}
\usepackage{amsfonts}
\usepackage{braket}
\usepackage[pdftex]{graphicx}
\usepackage{comment}

\begin{document}
\title {Predicted photoinduced pair annihilation of emergent magnetic charges in the organic salt $\alpha$-(BEDT-TTF)$_2$I$_3$ irradiated by linearly polarized light}
\author{Keisuke Kitayama}
\affiliation{Department of Physics, University of Tokyo, Hongo, Bunkyo-ku, Tokyo 113-8656, Japan}
\author{Masahito Mochizuki}
\affiliation{Department of Applied Physics, Waseda University, Okubo, Shinjuku-ku, Tokyo 169-8555, Japan}
\author{Yasuhiro Tanaka}
\affiliation{Department of Applied Physics, Waseda University, Okubo, Shinjuku-ku, Tokyo 169-8555, Japan}
\author{Masao Ogata}
\affiliation{Department of Physics, University of Tokyo, Hongo, Bunkyo-ku, Tokyo 113-8656, Japan}
\affiliation{Trans-scale Quantum Science Institute, University of Tokyo, Bunkyo-ku, Tokyo 113-0033, Japan}
\begin{abstract}
Prolonged experimental attempts to find magnetic monopoles (i.e., elementary particles with an isolated magnetic charge in three dimensions) have not yet been successful despite intensive efforts made since Dirac's proposal in 1931. Particle physicists have predicted the possible collision and pair annihilation of two magnetic charges with opposite signs. However, if such annihilation exists, its experimental observation would be difficult because its energy scale is predicted to be tremendously high ($\sim$10$^{16}$ GeV). In the present work, we theoretically predict using the Floquet theory that a pair of slightly gapped Dirac-cone bands in a weakly-charge-ordered organic conductor $\alpha$-(BEDT-TTF)$_2$I$_3$, which behave as magnetic charges with opposite signs in the momentum space, exhibit pair annihilation under irradiation with linearly polarized light. This photoinduced pair annihilation is accompanied by a non-topological phase transition to the Floquet normal insulator phase in contrast to the well-known circularly-polarized-light-induced topological phase transition to the Floquet Chern insulator phase. We discuss that $\alpha$-(BEDT-TTF)$_2$I$_3$ has a peculiar band structure capable of realizing a suitable experimental condition (i.e., off-resonant condition) and a charge ordered state providing a required staggered site potential and thereby provides a rare example of materials that can be used to observe the predicted pair annihilation phenomenon. The feasibility of experimental observation is also discussed.
\end{abstract}
%\pacs{}
\maketitle
%\sloppy 

\section{Introduction}
%%%%%%%%%%%%%%%%%%%%%%%%%%%%%%%
\begin{figure}[t]
\includegraphics[scale=0.5]{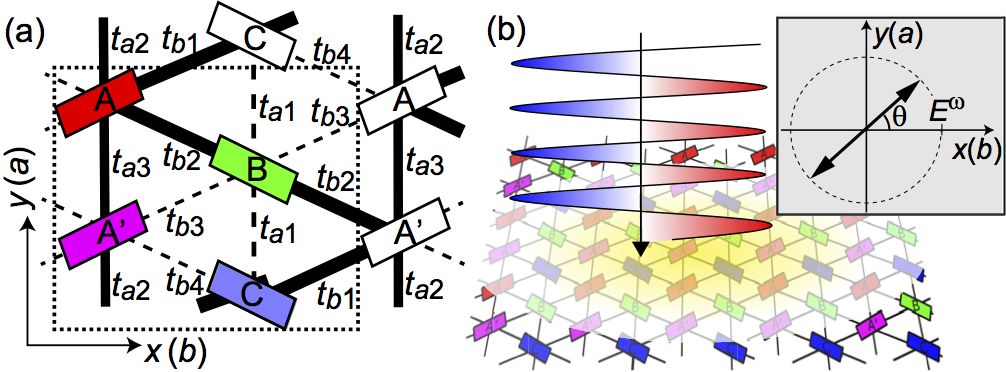}
\caption{(a) Crystal structure of the BEDT-TTF layer in $\alpha$-(BEDT-TTF)$_2$I$_3$ whose unit cell (dashed rectangle) contains four molecules (A, A$^\prime$, B and C). Transfer integrals for a tight-binding model of this compound are also shown. (b) Schematic illustration of $\alpha$-(BEDT-TTF)$_2$I$_3$ irradiated with linearly polarized light. The ac electric field of light is given in the form $-E^\omega \cos(\omega\tau)(\cos\theta, \sin\theta)$, where $E^\omega$ and $\theta$ are the amplitude and polarization angle.}
\label{Fig1}
\end{figure}
%%%%%%%%%%%%%%%%%%%%%%%%%%%%%%%%
In 1931, Paul Dirac proposed the concept of the magnetic monopole as an elementary particle with an isolated north or south magnetic pole from the viewpoint of electromagnetic duality in electromagnetism~\cite{Dirac31}. More than 40 years later, the magnetic monopole was found to be a topological soliton solution in non-Abelian gauge theory for grand unification~\cite{tHooft74,Polyakov74}. Despite intensive experimental efforts, magnetic monopoles have not yet been observed in nature. The experimental difficulty lies in the extremely heavy mass of the magnetic monopole, which was theoretically estimated to be 10$^{16}$ GeV~\cite{Polyakov74}.

Researchers have, however, discovered that gapped Dirac-cone bands in materials can behave as sources or sinks of an emergent magnetic field through exerting an additional quantum phase called the Berry phase on itinerant electrons~\cite{Berry84}. A source (sink) of the emergent magnetic field can be regarded as a positive (negative) magnetic charge, and the three-dimensional version is called a magnetic monopole (antimonopole). These emergent magnetic charges originating from the topology of the band structure often cause intriguing physical phenomena~\cite{Nagaosa12}. One important example is the anomalous Hall effect observed in the ferromagnetic perovskite SrRuO$_3$, in which the Berry curvature originating from its band topology behaves as a magnetic monopole in the momentum space; the resulting emergent magnetic field gives rise to a transverse electrical current by acting on the conduction electrons~\cite{FangZ03}. In addition to the specific band structure, real-space topological spin textures in magnets behave as emergent magnetic charges or magnetic monopoles by acting on electron spins via exchange interactions~\cite{Volovik87,Kotiuga89,Castelnovo08,Morris09,Milde13,Kanazawa16}. The resulting emergent magnetic fields of spin topology origin induce a Hall effect of conduction electrons, called the topological Hall effect. Importantly, the magnetic charges in these condensed-matter systems emerge in an accessible energy range of approximately 0.1-1 eV. These systems therefore provide unique playgrounds for studying the electromagnetic properties and phenomena of magnetic charges and magnetic monopoles.

Pair annihilation is an important phenomenon associated with magnetic charges or magnetic monopoles. However, its experimental observation is extremely difficult for real monopoles and antimonopoles predicted in the grand unified theory~\cite{tHooft74,Polyakov74}, if they exist at all, because of the heavy masses of the monopoles and antimonopoles. Recent experiments have revealed the pair annihilation of emergent monopoles and antimonopoles in the cubic chiral-lattice magnet MnGe under the application of a magnetic field~\cite{Kanazawa16}. In this magnet, monopoles and antimonopoles simultaneously exist as hedgehog-type topological spin textures when there is no magnetic field. When a magnetic field is applied, the distances between the monopoles and antimonopoles decrease because of the field-induced modulation of the spin textures, and their pair annihilation eventually occurs. This phenomenon is a field-induced pair annihilation of magnetic charges of spin topology origin. On the other hand, the pair annihilation originating from band topology has not yet been observed.

In this paper, we theoretically propose that the pair annihilation of emergent magnetic charges with opposite signs is realized in photodriven organic conductor $\alpha$-(BEDT-TTF)$_2$I$_3$, where BEDT-TTF denotes bis(ethylenedithio)-tetrathiafulvalene~\cite{Tajima06}. This compound possesses a pair of tilted Dirac cones in the band structure~\cite{Katayama06,Kobayashi07,Tajima09,Osada08,Konoike12,Hirata16,Kajita14}, which can have a small gap at the Dirac points when there is a staggered site potential due to a weak charge order~\cite{Tajima06}. The gapped Dirac cones behave as magnetic charges with opposite signs in the two-dimensional momentum space. Using the Floquet theory~\cite{Floquet,Mikami16}, we demonstrate that their pair annihilation occurs under irradiation with linearly polarized light in the energy regime of less than 1 eV [see Fig.~\ref{Fig1}]. Note that this phenomenon is distinct from the photoinduced topological phase transition induced by circularly polarized light, which has been predicted and discussed in recent theoretical works~\cite{Kitayama20,Tanaka21}. The pair annihilation occurs under linearly polarized light, which does not break the time-reversal symmetry, whereas the broken time-reversal symmetry caused by circularly polarized light is required for the photoinduced topological phase transition. Since the emergent magnetic charges in this organic system appear in the two-dimensional momentum space, they can be regarded as emergent magnetic fluxes rather than magnetic monopoles in the momentum space. To date, the pair annihilation of antiparallel magnetic fluxes has been observed in a superconductor by a real-space imaging technique using the Lorentz transmission electron microscopy~\cite{Tonomura05}. The photoinduced pair annihilation of magnetic charges predicted here can be regarded as an analog of this phenomenon with emergent magnetic fluxes of electronic band topology origin. Our proposal may provide a unique opportunity to study the physics of the pair annihilation of magnetic charges by measurements with small-scale equipment instead of large experimental facilities.

\section{Model and Methods}
To investigate a photodriven organic conductor $\alpha$-(BEDT-TTF)$_2$I$_3$ irradiated with linearly polarized light, we start with a tight-binding model that describes the electronic structure in the BEDT-TTF layer of the compound before light irradiation~\cite{Katayama06,Kajita14}:
%%%%%%%%%%%%%%%%%%%%%%%%%
\begin{eqnarray}
H = \sum_{\langle i,j \rangle}\sum_{\alpha, \beta} 
t_{i\alpha,j\beta}c_{i,\alpha}^{\dagger}c_{j,\beta}
+\Delta \sum_i (c_{i,{\rm A}}^{\dagger}c_{i,{\rm A}}-c_{i,{\rm A}^\prime}^{\dagger}c_{i,{\rm A}^\prime}).
\label{eq:TBm1}
\end{eqnarray}
%%%%%%%%%%%%%%%%%%%%%%%%%
Here, $i$ and $j$ denote the unit cells whereas $\alpha$ and $\beta$ denote the molecular sites (A, A$^\prime$, B and C). The symbol $c_{j,\beta}^{\dagger}$ ($c_{i,\alpha}$) denotes the electron creation (annihilation) operator whereas $t_{i\alpha,j\beta}$ denotes transfer integrals between neighboring sites. This compound exhibits a charge order with a band gap, whose transition temperature is $T_{\rm CO}=135$ K at ambient pressure~\cite{Tajima06}. This charge order melts under a uniaxial pressure ($P_a>4$ kbar) along the $a$ axis. Around and above the threshold pressure of $P_a \sim 4$ kbar, a zero-gap semiconducting state with a pair of Dirac cones between the third and fourth bands appears. The Fermi level lies between these two bands because the electron filling of this compound is 3/4. In Eq.~(\ref{eq:TBm1}), we introduce a staggered site potential ($\Delta>0$) that corresponds to the order parameter for the charge order in a mean-field theory~\cite{Seo00}. When the charge order is absent ($\Delta=0$), the sites A and A$^\prime$ are equivalent to each other owing to inversion symmetry. Meanwhile, in the charge-ordered phase ($\Delta>0$), the site A (A$^\prime$) becomes hole rich (poor), as observed in various experiments such as Raman-spectroscopy measurements~\cite{Wojciechowski03}. For nonzero $\Delta$, the gaps form at the Dirac points, leading to finite peaks in the Berry curvature instead of their divergence as in the case of $\Delta=0$. The finite peaks in the Berry curvature can be interpreted as emergent magnetic charges. 
%It is noted that although the relativistic spin--orbit interaction is another possible origin of the gap opening at the Dirac points~\cite{Winter17,Osada18}, the contribution of the staggered site potential dominates the contribution of the relativistic spin--orbit interaction in the presence of the charge order because the interaction is rather weak in the present organic compound.

The magnitude of $\Delta$ can be estimated as $\Delta \sim k_{\mathrm{B}} T_{\mathrm{CO}}$ using the mean-field theory. We have $T_{\mathrm{CO}}=135$ K for $\alpha$-(BEDT-TTF)$_2$I$_3$, and the magnitude of $\Delta$ can thus be tuned within a range of $0\leq\Delta\lesssim 0.01$ eV by applying uniaxial pressure or raising the temperature through varying the extent of charge disproportionation. In the present study, most of the calculations are performed by setting $\Delta$ = 0 as a limit case of the weak charge order. Meanwhile, the Berry curvatures of the bands are calculated by introducing negligibly small site potentials of $\Delta=1\times 10^{-5}$ eV, with which the Dirac points become weakly gapped, and thereby the Berry curvatures are well defined. These small site potentials can be realized in a weakly-charge-ordered system located near the phase boundary with the zero-gap semiconducting state. 

The transfer integrals under uniaxial pressure are given by the following formulae deduced theoretically~\cite{Kobayashi04}: $t_{a1} = -0.028(1 + 0.089P_a)$, $t_{a2} = 0.048(1 + 0.167P_a)$, $t_{a3} = -0.020(1 - 0.025P_a)$, $t_{b1} = 0.123$, $t_{b2} = 0.140(1 + 0.011P_a)$, $t_{b3} = -0.062(1 + 0.032P_a)$ and $t_{b4} = -0.025$ (in units of eV)~\cite{Kobayashi04}. We assume that $P_a=4$ kbar in the following calculations.
%%%%%%%%%%%%%%%%%%%%%%%%%
%\begin{align}
%&t_{a1} = -0.028(1 + 0.089P_a) \;{\rm eV},\nonumber \\
%&t_{a2} = 0.048(1 + 0.167P_a) \;{\rm eV},\nonumber \\
%&t_{a3} = -0.020(1 - 0.025P_a) \;{\rm eV},\nonumber \\
%&t_{b1} = 0.123 \;{\rm eV},\nonumber \\
%&t_{b2} = 0.140(1 + 0.011P_a) \;{\rm eV},\nonumber \\
%&t_{b3} = -0.062(1 + 0.032P_a) \;{\rm eV},\nonumber \\
%&t_{b4} = -0.025 \;{\rm eV}.\nonumber
%\end{align}
%%%%%%%%%%%%%%%%%%%%%%%%%
We introduce a time-dependent vector potential,
%%%%%%%%%%%%%%%%%%%%%%%%%
\begin{eqnarray}
\bm A(\tau)=A\sin(\omega\tau)(\cos \theta, \sin \theta),
\label{eq:vecpot}
\end{eqnarray}
%%%%%%%%%%%%%%%%%%%%%%%%%
where $\omega$ is the angular frequency. This vector potential produces a linearly polarized light electric field whose polarization angle is $\theta$ [see the inset of Fig.~\ref{Fig1}(b)],
%%%%%%%%%%%%%%%%%%%%%%%%%
\begin{eqnarray}
\bm E(\tau)=-\frac{d\bm A(\tau)}{d\tau}=-E^\omega\cos(\omega\tau)(\cos \theta, \sin \theta). 
\label{eq:acEfld}
\end{eqnarray}
%%%%%%%%%%%%%%%%%%%%%%%%%
Here we define the amplitude of light $E^\omega$ as $E^\omega = A\omega$. The effect of light irradiation is considered by attaching the Peierls phase to the transfer integrals in Eq.~(\ref{eq:TBm1}). We obtain a time-dependent tight-binding Hamiltonian $H(\tau)$ for the photodriven system~\cite{Yonemitsu06,Kitayama20,Tanaka10,Miyashita10}:
%%%%%%%%%%%%%%%%%%%%%%%%%
\begin{eqnarray}
H(\tau)&=&\sum_{\langle i,j \rangle}\sum_{\alpha, \beta} t_{i\alpha,j\beta} \;
e^{-ie\bm{A}(\tau)\cdot(\bm{r}_{i\alpha} - \bm{r}_{j\beta})/\hbar}
c_{i,\alpha}^{\dagger}c_{j,\beta}
\nonumber \\
&+&\Delta \sum_i (c_{i,{\rm A}}^{\dagger}c_{i,{\rm A}}-c_{i,{\rm A}^\prime}^{\dagger}c_{i,{\rm A}^\prime}),
\label{eq:TBm2}
\end{eqnarray}
%%%%%%%%%%%%%%%%%%%%%%%%%
%%%%%%%%%%%%%%%%%%%%%%%%%
%\begin{eqnarray}
%H(\tau)=\sum_{\langle i,j \rangle}\sum_{\alpha, \beta} t_{i\alpha,j\beta} \;
%\exp\left\{-i\frac{e}{\hbar}\bm{A}(\tau)\cdot(\bm{r}_{i\alpha} - \bm{r}_{j\beta})\right\}
%c_{i,\alpha}^{\dagger}c_{j,\beta},
%\label{eq:TBm2}
%\end{eqnarray}
%%%%%%%%%%%%%%%%%%%%%%%%%
where $\bm r_{i\alpha}=(b\tilde{x}_{i\alpha}, a\tilde{y}_{i\alpha})$ denotes the coordinates of the $\alpha$th molecular sublattice in the $i$th unit cell. For the lattice constants along the $y$ and $x$ axes, we use experimentally measured values of $a$=0.9187 nm and $b$=1.0793 nm, respectively~\cite{Mori12}.

Time evolutions of the wavefunction in the light-irradiated $\alpha$-(BEDT-TTF)$_2$I$_3$ are described by the time-dependent Schr\"{o}dinger equation,
%%%%%%%%%%%%%%%%%%%%%%%%%
\begin{eqnarray}
i \hbar \partial_\tau \ket{\Psi(\tau)} = H(\tau)\ket{\Psi(\tau)}. 
\label{eq:Scheq}
\end{eqnarray}
%%%%%%%%%%%%%%%%%%%%%%%%%
The Hamiltonian $H(\tau)$ is time periodic [i.e., $H(\tau)=H(\tau+T)$, with $T$ being the temporal periodicity of the light field], and therefore we can apply the Floquet theorem to this equation. This theorem can be regarded as a temporal version of the Bloch theorem~\cite{Floquet,Mikami16}. According to the Floquet theorem, the wavefunction $\ket{\Psi(\tau)}$ is written in the form
%%%%%%%%%%%%%%%%%%%%%%%%%
\begin{eqnarray}
\ket{\Psi(\tau)} = e^{i\varepsilon \tau/\hbar}\ket{\Phi(\tau)},
\label{eq:Flqthm}
\end{eqnarray}
%%%%%%%%%%%%%%%%%%%%%%%%%
where $\ket{\Phi(\tau)}$ is referred to as the Floquet state that satisfies $\ket{\Phi(\tau)}=\ket{\Phi(\tau +T)}$, whereas $\varepsilon$ is its quasienergy. We substitute Eq.~(\ref{eq:Flqthm}) into Eq.~(\ref{eq:Scheq}) and perform the Fourier transformations with respect to time. We obtain
%%%%%%%%%%%%%%%%%%%%%%%%%
\begin{eqnarray}
\sum_{m=-\infty}^{\infty} \mathcal{H}_{nm} \ket{\Phi_{\nu}^m}
=\varepsilon^n_{\nu}\ket{\Phi_{\nu}^n},
\label{eq:H-Mw1}
\end{eqnarray}
%%%%%%%%%%%%%%%%%%%%%%%%%
with
%%%%%%%%%%%%%%%%%%%%%%%%%
\begin{eqnarray}
\mathcal{H}_{nm}=H_{n-m}-m\omega\delta_{n,m}.
\label{eq:H-Mw2}
\end{eqnarray}
%%%%%%%%%%%%%%%%%%%%%%%%%
Here $n$ and $m$ correspond to the number of photons, and $\nu$ labels the eigenstates in each subspace of the photon number. The matrix $\mathcal{H}$ is referred to as the Floquet matrix. The Fourier components $H_n$ and $\ket{\Phi_{\nu}^n}$ are defined by
%%%%%%%%%%%%%%%%%%%%%%%%%
\begin{eqnarray}
\ket{\Phi_{\nu}^n}&=&\frac{1}{T}\int_0^T \ket{\Phi_{\nu}(\tau)}e^{in\omega \tau} d\tau,
\label{phiFC}\\
H_n&=&\frac{1}{T}\int_0^T H(\tau)e^{in\omega \tau} d\tau.
\label{HamFC}
\end{eqnarray}
%%%%%%%%%%%%%%%%%%%%%%%%%

Importantly, the eigenvalue equation in Eq.~(\ref{eq:H-Mw1}) is no longer time dependent. The problem of the nonequilibrium steady states in the time-periodically driven system is effectively mapped onto a problem of the equilibrium states. For practical calculations, we truncate the Floquet matrix of originally infinite dimension and consider a matrix of finite dimension of $|m|\leq 8$. The physical meaning of this truncation is that the processes with absorption and emission of a large number ($|m|<8$) of photons are neglected. The number of photons $m$ to be considered is determined by the ratio between the bandwidth $W$ and the light frequency $\hbar\omega$ as $m>W/(\hbar\omega)$~\cite{Mikami16}. Conversely, the truncated Floquet matrix of $|m| \leq 8$ provides accurate results when $\hbar \omega \gtrsim 0.1$ eV because the typical bandwidth of $\alpha$-(BEDT-TTF)$_2$I$_3$ is $W \sim 0.8$ eV.

In addition to directly solving the eigenequation Eq.~(\ref{eq:H-Mw1}), we examine another approach based on the effective Hamiltonian in the high frequency limit, which is derived using the Brillouin--Wigner theorem~\cite{Mikami16},
%%%%%%%%%%%%%%%%%%%%%%%%%
\begin{eqnarray}
H_{\mathrm{eff}}=H_0
+\sum_{n=1}^\infty \frac{[H_{-n}, H_n]}{n\hbar\omega}
+O\left(\frac{W^3}{\hbar^2\omega^2}\right),
\label{eq:Heff}
\end{eqnarray}
%%%%%%%%%%%%%%%%%%%%%%%%%
where $H_n$ is defined in Eq.~(\ref{HamFC}), and $H_0$ is the 0th Fourier coefficient $H_{n=0}$. The index $+n$ ($-n$) denotes the number of emitted (absorbed) photons. Using the time-periodic tight-binding Hamiltonian in Eq.~(\ref{eq:TBm2}) for photodriven $\alpha$-(BEDT-TTF)$_2$I$_3$, we obtain
%%%%%%%%%%%%%%%%%%%%%%%%%
\begin{eqnarray}
H_n &=& \sum_{\langle i,j \rangle}\sum_{\alpha, \beta} t_{i\alpha,j\beta} 
J_n(A_{i\alpha,j\beta})c_{i,\alpha}^{\dagger}c_{j,\beta}
\nonumber \\
&+&\Delta \sum_i (c_{i,{\rm A}}^{\dagger}c_{i,{\rm A}}-c_{i,{\rm A}^\prime}^{\dagger}c_{i,{\rm A}^\prime}),
\label{eq:HnLPL}
\end{eqnarray}
%%%%%%%%%%%%%%%%%%%%%%%%%
where
%%%%%%%%%%%%%%%%%%%%%%%%%
\begin{eqnarray}
A_{i\alpha,j\beta}
&=&\frac{eA}{\hbar}
\left[b\cos\theta(\tilde{x}_{i\alpha}-\tilde{x}_{j\beta})
     +a\sin\theta(\tilde{y}_{i\alpha}-\tilde{y}_{j\beta}) \right]
\nonumber \\
&=&
 \mathcal{A}_b(\tilde{x}_{i\alpha}-\tilde{x}_{j\beta})
+\mathcal{A}_a(\tilde{y}_{i\alpha}-\tilde{y}_{j\beta}),
\label{eq:Aij1}
\end{eqnarray}
%%%%%%%%%%%%%%%%%%%%%%%%%
with
%%%%%%%%%%%%%%%%%%%%%%%%%
\begin{eqnarray}
\mathcal{A}_a&=&eaA\sin\theta/\hbar,
\quad\quad
 \mathcal{A}_b=ebA\cos\theta/\hbar.
\label{eq:Aij2}
\end{eqnarray}
%%%%%%%%%%%%%%%%%%%%%%%%%
Here, $J_n$ is the $n$th Bessel function.

After conducting Fourier transformations with respect to the spatial coordinates, we obtain
%%%%%%%%%%%%%%%%%%%%%%%%%
\begin{eqnarray}
\hat{H}_n(\bm{k})=\left(
\begin{array}{cccc}
\Delta\delta_{n,0} & A_{2,n} & B_{2,n} & B_{1,n} \\
A_{2,-n}^{*} & -\Delta\delta_{n,0} & B_{2,-n}^{*} & B_{1,-n}^{*} \\
B_{2,-n}^{*} & B_{2,n} & 0 & A_{1,n} \\
B_{1,-n}^{*} & B_{1,n} & A_{1,n} & 0
\end{array}
\right),
\label{eq:Hk}
\end{eqnarray}
%%%%%%%%%%%%%%%%%%%%%%%%%
where
%%%%%%%%%%%%%%%%%%%%%%%%%
\begin{align}
&A_{1,n}=
 t_{a1}\,e^{ i\frac{k_y}{2}}J_{-n}(\mathcal{A}_a/2)
+t_{a1}\,e^{-i\frac{k_y}{2}}J_{n}(\mathcal{A}_a/2),
\nonumber \\
&A_{2,n}=
 t_{a2}\,e^{ i\frac{k_y}{2}}J_{-n}(\mathcal{A}_a/2)
+t_{a3}\,e^{-i\frac{k_y}{2}}J_{n}(\mathcal{A}_a/2),
\nonumber \\
&B_{1,n}=
 t_{b1}\,e^{ i(\frac{k_x}{2}+\frac{k_y}{4})}J_{-n}(\mathcal{A}_+)
+t_{b4}\,e^{-i(\frac{k_x}{2}-\frac{k_y}{4})}J_{-n}(\mathcal{A}_-),
\nonumber \\
&B_{2,n}=
 t_{b2}\,e^{ i(\frac{k_x}{2}-\frac{k_y}{4})}J_{n}(\mathcal{A}_-)
+t_{b3}\,e^{-i(\frac{k_x}{2}+\frac{k_y}{4})}J_{n}(\mathcal{A}_+),
\nonumber
\end{align}
%%%%%%%%%%%%%%%%%%%%%%%%%
with
%%%%%%%%%%%%%%%%%%%%%%%%%
\begin{eqnarray}
\mathcal{A}_{\pm}&=&\frac{eA}{4\hbar}(\pm2b\cos\theta + a\sin\theta).
\label{eq:Aij3}
\end{eqnarray}
%%%%%%%%%%%%%%%%%%%%%%%%%
In the following calculations, we use Eq.~(\ref{eq:Hk}) for $H_n$ in Eqs.~(\ref{eq:H-Mw2}) and (\ref{eq:Heff}).

\section{Results}
%%%%%%%%%%%%%%%%%%%%%%%%%%%%%%%
\begin{figure*} %[htb]
\includegraphics[scale=0.5]{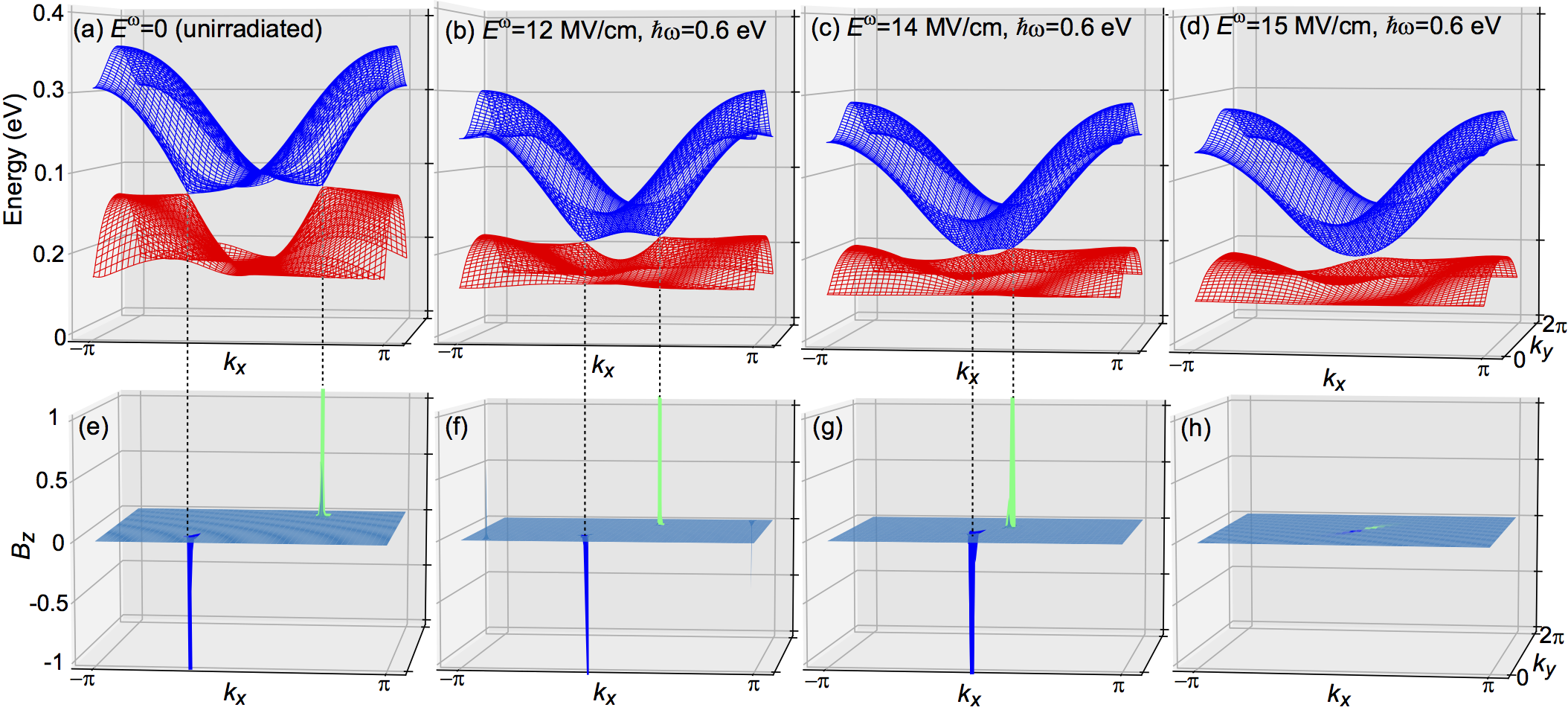}
\caption{(a--d) Quasienergy band structures of the photodriven $\alpha$-(BEDT-TTF)$_2$I$_3$ under irradiation with linearly polarized light for various light amplitudes $E^\omega$: (a) $E^\omega$=0 (unirradiated case), (b) $E^\omega$=12 MV/cm, (c) $E^\omega$=14 MV/cm and (d) $E^\omega$=15 MV/cm. (e--f) Berry curvature of the fourth band in the photodriven $\alpha$-(BEDT-TTF)$_2$I$_3$. The sharp peaks with opposite signs indicate the existence of positive and negative magnetic charges at the momentum points corresponding to the gapped Dirac points. The frequency and the polarization angle of light are fixed at $\hbar\omega = 0.6$ eV and $\theta=45^\circ$, respectively. The pair annihilation of magnetic charges is observed with increasing $E^\omega$.}
\label{Fig2}
\end{figure*}
%%%%%%%%%%%%%%%%%%%%%%%%%%%%%%%%
We first calculate quasienergy band structures of the photodriven $\alpha$-(BEDT-TTF)$_2$I$_3$ by diagonalizing the Floquet Hamiltonian matrix in Eq.~(\ref{eq:H-Mw2}). Figures~\ref{Fig2}(a)--(d) show the obtained band structures for various light amplitudes $E^\omega$. Here, the frequency and polarization angle of the linearly polarized light are fixed at $\hbar\omega = 0.6$ eV and $\theta=45^\circ$, respectively. Figure~\ref{Fig2}(a) shows the band structure before the light irradiation; this band structure has a pair of Dirac cones. Meanwhile, Figs.~\ref{Fig2}(b)--(d) show that the band structures during light irradiation have a photoinduced pair annihilation of the Dirac points. The distance between the two Dirac points shortens as the light amplitude $E^\omega$ increases, and these two Dirac points eventually merge and disappear at $E^\omega \sim 15$ MV/cm.

This phenomenon can be regarded as a pair annihilation of emergent magnetic charges with opposite signs; that is, a sink and source of the emergent magnetic field. This pair annihilation in the momentum space is clearly visualized by the Berry curvature in the momentum space, which acts as an effective magnetic field by exerting an additional quantum phase (Berry phase) on itinerant electrons. The Berry curvature $B_z^{n\nu}(\bm k)$ of the $\nu$th band in the $n$-photon subspace is calculated using the formula~\cite{Berry84}
%%%%%%%%%%%%%%%%%%%%%%%%%%%%%%%%%%
\begin{align}
&B_z^{n\nu}(\bm k)=
\nonumber \\
&i\sum_{(m,\mu)}\frac{
\bra{\Phi_{\nu}^n(\bm k)}\frac{\partial \hat{\mathcal{H}}}{\partial k_x}\ket{\Phi_{\mu}^m(\bm k)}
\bra{\Phi_{\mu}^m(\bm k)}\frac{\partial \hat{\mathcal{H}}}{\partial k_y}\ket{\Phi_{\nu}^n(\bm k)}
-{\rm c.c.}}
{[\varepsilon^m_\mu(\bm k)-\varepsilon^n_\nu(\bm k)]^2}.
\end{align}
%%%%%%%%%%%%%%%%%%%%%%%%%%%%%%%%%%
Here, $\hat{\mathcal{H}}$ denotes the Floquet Hamiltonian matrix, whereas $\varepsilon^n_\nu(\bm k)$ and $\ket{\Phi_{\nu}^n(\bm k)}$ are the eigenenergies and eigenvectors of Eq.~(\ref{eq:H-Mw1}) with $\nu=1,2,3,4$ and $|n|\le 8$. The summation is taken over $m$ and $\mu$, where $(m,\mu)\ne(n,\nu)$; ${\rm c.c.}$ denotes the complex conjugate of the first term in the numerator. For the massless Dirac-cone bands without a gap opening, the Berry curvature diverges at the Dirac points. Here, the tiny gap at the Dirac points due to the staggered site potential of $\Delta = 1 \times 10^{-5}$ eV is assumed to suppress this divergence. Figures~\ref{Fig2}(e)--(h) present the calculated Berry curvatures $B_z^{04}(\bm k)$ of the fourth band ($\nu$=4) in the zero-photon subspace ($n$=0) for various light amplitudes $E^\omega$, which respectively correspond to the band structures in Figs.~\ref{Fig2}(a)--(d). Note that the peaks of $B_z^{n\nu}(\bm k)$ at the two gapped Dirac points have opposite signs~\cite{Suzumura11,Osada17,Kitayama20}. The positive peak corresponds to a positive magnetic charge (i.e., a source of the emergent magnetic field) whereas the negative peak corresponds to a negative magnetic charge (i.e., a sink of the emergent magnetic field). These magnetic charges can be regarded as magnetic fluxes rather than magnetic monopoles because they emerge in the two-dimensional momentum space. Therefore, the pair annihilation in this organic system can be interpreted as the pair annihilation of magnetic fluxes. 

This pair annihilation of emergent magnetic charges in the momentum space is accompanied by a nonequilibrium phase transition. To study the photoinduced phases in the present material, we define two types of energy gap:
%%%%%%%%%%%%%%%%%%%%%%%%%%%%%%%
\begin{align}
&E_{\rm gap}=\min[\varepsilon^0_4(\bm k)]- \max[\varepsilon^0_3(\bm k)],
\label{eq:Egap1}
\\
&\tilde{E}_{\rm gap}=\min[\varepsilon^0_4(\bm k) - \varepsilon^0_3(\bm k)].
\label{eq:Egap2}
\end{align}
%%%%%%%%%%%%%%%%%%%%%%%%%%%%%%%
The gap $E_{\rm gap}$ is an indicator used to judge whether the system is insulating. In other words, when $E_{\rm gap}>0$, a gap opens at the Fermi level over the whole area of the Brillouin zone, and the system is thus insulating. In contrast, the system is semimetallic when $E_{\rm gap}<0$. Meanwhile, $\tilde{E}_{\rm gap}$ is an indicator used to judge whether the Dirac cones are gapped. That is to say, the Dirac cones are gapped at the Dirac points when $\tilde{E}_{\rm gap}>0$, whereas they are not gapped when $\tilde{E}_{\rm gap}=0$. Note that $\tilde{E}_{\rm gap}$ cannot be negative by definition.

%%%%%%%%%%%%%%%%%%%%%%%%%%%%%%%
\begin{figure*} %[htb]
\includegraphics[scale=0.5]{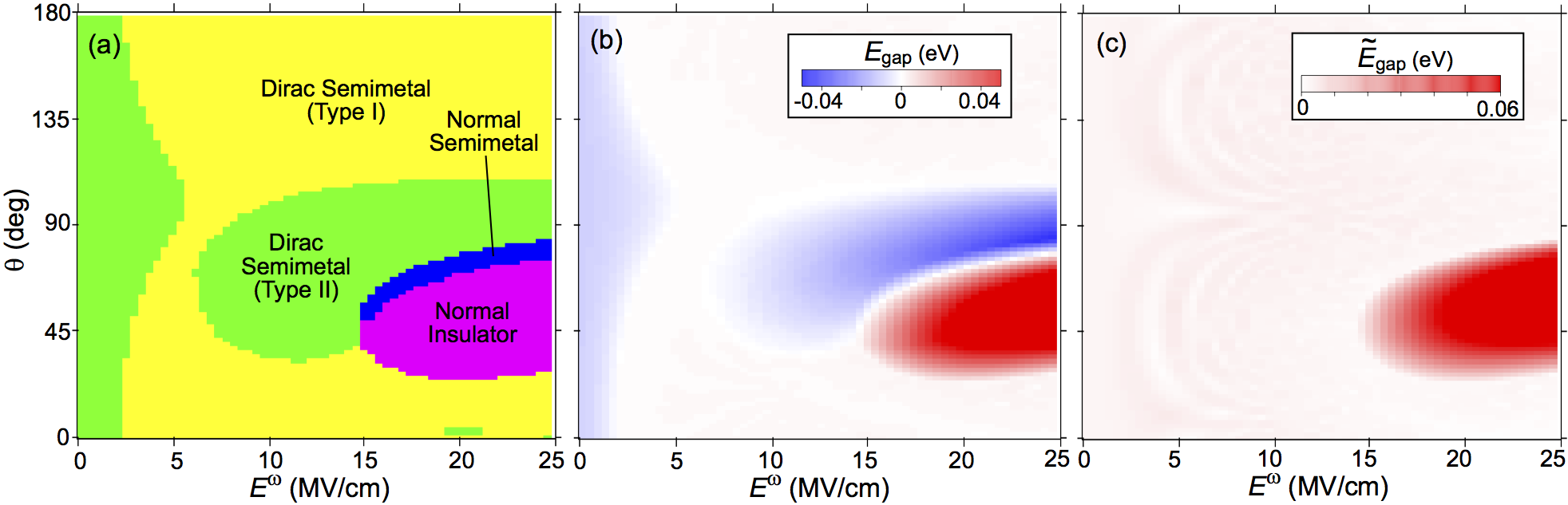}
\caption{(a) Phase diagram for nonequilibrium steady states in the photodriven $\alpha$-(BEDT-TTF)$_2$I$_3$ under irradiation with linearly polarized light, in the plane of the amplitude $E^\omega$ and the polarization angle $\theta$ of the light. (b)--(c) Color maps of the calculated two types of energy gap, (b) $E_{\rm gap}$ and (c) $\tilde{E}_{\rm gap}$, defined in Eqs. (\ref{eq:Egap1}) and (\ref{eq:Egap2}), respectively. The light frequency $\omega$ is fixed at $\hbar\omega$=0.6 eV.}
\label{Fig3}
\end{figure*}
\begin{figure} %[htb]
\includegraphics[scale=0.5]{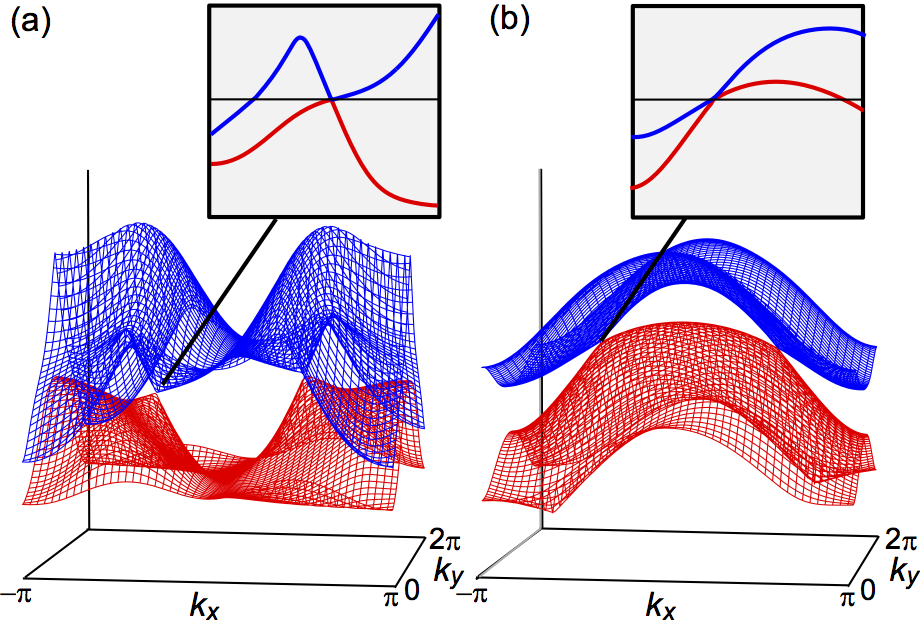}
\caption{Schematic illustrations of two possible types of band structure for $E_{\rm gap}<0$ and $\tilde{E}_{\rm gap}=0$, both of which are expected to behave as a semimetal. (a) Band structure in which the upper band crosses the Fermi level at momenta far from the Dirac cones. (b) Band structure with overtilted Dirac cones where the upper (lower) cone band is located below (above) the Fermi level. Both cases are assigned to the type-II Dirac semimetal phase in the phase diagram in Fig.~\ref{Fig3}(a).}
\label{Fig4}
\end{figure}

%%%%%%%%%%%%%%%%%%%%%%%%%%%%%%%%
%%%%%%%%%%%%%%%%%%%%%%%%%%%%%%%
\begin{table}[tb]
\begin{tabular}{|c||c|c|}
\hline
         & $\tilde{E}_{\rm gap}=0$ & $\tilde{E}_{\rm gap}>0$ \\
\hline \hline
$E_{\rm gap}>0$ & Type I Dirac semimetal & Normal insulator \\ 
\hline
$E_{\rm gap}\le 0$ & Type II Dirac semimetal & Normal semimetal \\ 
\hline
\end{tabular}
\caption{Classification of the photoinduced phases in $\alpha$-(BEDT-TTF)$_2$I$_3$ irradiated with linearly polarized light according to the band gaps $E_{\rm gap}$ and $\tilde{E}_{\rm gap}$ defined by Eqs.~(\ref{eq:Egap1}) and (\ref{eq:Egap2}).}
\label{tab:phsd}
\end{table}
%%%%%%%%%%%%%%%%%%%%%%%%%%%%%%%
Figure~\ref{Fig3}(a) presents the phase diagram for nonequilibrium steady states in $\alpha$-(BEDT-TTF)$_2$I$_3$ irradiated with linearly polarized light in the plane of the amplitude $E^\omega$ and the polarization angle $\theta$ of light. Here, the light frequency $\omega$ is fixed at $\hbar\omega$=0.6 eV. This phase diagram contains a variety of nonequilibrium steady phases; for example, the type-I Dirac semimetal, type-II Dirac semimetal, normal semimetal and normal insulator phases. These phases are classified according to the signs of the calculated energy gaps $E_{\rm gap}$ [Fig.~\ref{Fig3}(b)] and $\tilde{E}_{\rm gap}$ [Fig.~\ref{Fig3}(c)] (see also Table~\ref{tab:phsd}). Note that when ${E}_{\rm gap}<0$ and $\tilde{E}_{\rm gap}=0$, two different types of band structure are possible [see Figs~\ref{Fig4}(a) and (b)]. It is noted that we did not consider the staggered site potential in obtaining Fig.~\ref{Fig3}.

%%%%%%%%%%%%%%%%%%%%%%%%%%%%%%%
%%%%%%%%%%%%%%%%%%%%%%%%%%%%%%%%
Pair annihilation occurs at $E^\omega$=15 MV/cm in the case that $\theta$=45$^\circ$ [see Fig.~\ref{Fig2}]. As shown in Fig.~\ref{Fig3}(a), the phase transition from the Dirac semimetal phase to the normal insulator phase occurs at $E^\omega$=15 MV/cm, which indicates that the observed pair annihilation of magnetic charges is accompanied by this photoinduced phase transition. Importantly, this phase transition is not topological but topologically trivial. In our previous work in Ref.~\cite{Kitayama20}, we argued that the Floquet topological insulator phase with a quantized nonzero Chern number emerges in the photodriven $\alpha$-(BEDT-TTF)$_2$I$_3$ under irradiation with circularly polarized light. In that case, the circularly polarized light breaks the time reversal symmetry, and the second term of the Floquet effective Hamiltonian in Eq.~(\ref{eq:Heff}) thus becomes finite ($\sum_{n=1}^\infty [H_{-n},H_n]/n\hbar\omega \ne 0$).
%%%%%%%%%%%%%%%%%%%%%%%%%%%%%%%%
%\begin{eqnarray}
%\sum_{n=1}^\infty \frac{[H_{-n},H_n]}{n\hbar\omega} \ne 0.
%\end{eqnarray}
%%%%%%%%%%%%%%%%%%%%%%%%%%%%%%%%
Indeed, this term opens a gap at the Dirac points, and the Floquet Chern insulator phase eventually appears as bands separated by the gap attain nonzero Chern numbers. Meanwhile, the Chern insulator phase never appears in the present case because the linearly polarized light does not break the time reversal symmetry. When the system is time-reversal invariant, the second term of the Floquet effective Hamiltonian vanishes ($\sum_{n=1}^\infty [H_{-n},H_n]/n\hbar\omega=0$) because the Hamiltonian is required to be invariant upon the replacement of $\omega$ with $-\omega$. Consequently, the Hamiltonian reduces to $H_{\rm eff}$=$H_0+\mathcal{O}(1/\omega^2)$. However, the photoinduced gap opening occurs in the present case because of the photoinduced renormalization of the transfer integrals. More specifically, according to Eq.~(\ref{eq:HnLPL}), the transfer integrals in the expression for $H_0$ are renormalized as
%%%%%%%%%%%%%%%%%%%%%%%%%%%%%%%%
\begin{eqnarray}
t_{i\alpha,j\beta} \;\rightarrow\; t_{i\alpha,j\beta}J_0(A_{i\alpha,j\beta}).
\label{renorm}
\end{eqnarray}
%%%%%%%%%%%%%%%%%%%%%%%%%%%%%%%%
The normal insulator phase appears when the gap is opened by the resulting photoinduced band deformation.

%%%%%%%%%%%%%%%%%%%%%%%%%%%%%%%%
\begin{table}[tbh]
\begin{tabular}{|c|c|c|}
\hline
Bond      & Photoinduced renormalization & Renormalization \\
direction & of transfer integral & factor \\
\hline
[010] & $t_{a1}$\;$\rightarrow$\;$t_{a1}J_0(\mathcal{A}_a/2)$ & $J_0(\mathcal{A}_a/2)$ \\
      & $t_{a2}$\;$\rightarrow$\;$t_{a2}J_0(\mathcal{A}_a/2)$ & \\
      & $t_{a3}$\;$\rightarrow$\;$t_{a3}J_0(\mathcal{A}_a/2)$ & \\
\hline
[110] & $t_{b1}$\;$\rightarrow$\;$t_{b1}J_0(\mathcal{A}_+)$ & $J_0(\mathcal{A}_+)$ \\
      & $t_{b3}$\;$\rightarrow$\;$t_{b3}J_0(\mathcal{A}_+)$ & \\
\hline
[1$\bar{1}$0] & $t_{b2}$\;$\rightarrow$\;$t_{b2}J_0(\mathcal{A}_-)$ & $J_0(\mathcal{A}_-)$ \\
              & $t_{b4}$\;$\rightarrow$\;$t_{b4}J_0(\mathcal{A}_-)$ & \\
\hline
\end{tabular}
\caption{Photoinduced anisotropic renormalizations of the transfer integrals in $\alpha$-(BEDT-TTF)$_2$I$_3$ under irradiation with linearly polarized light.}
\label{tab:trenorm}
\end{table}
%%%%%%%%%%%%%%%%%%%%%%%%%%%%%%%%
%%%%%%%%%%%%%%%%%%%%%%%%%%%%%%%
\begin{figure} %[htb]
\includegraphics[scale=0.5]{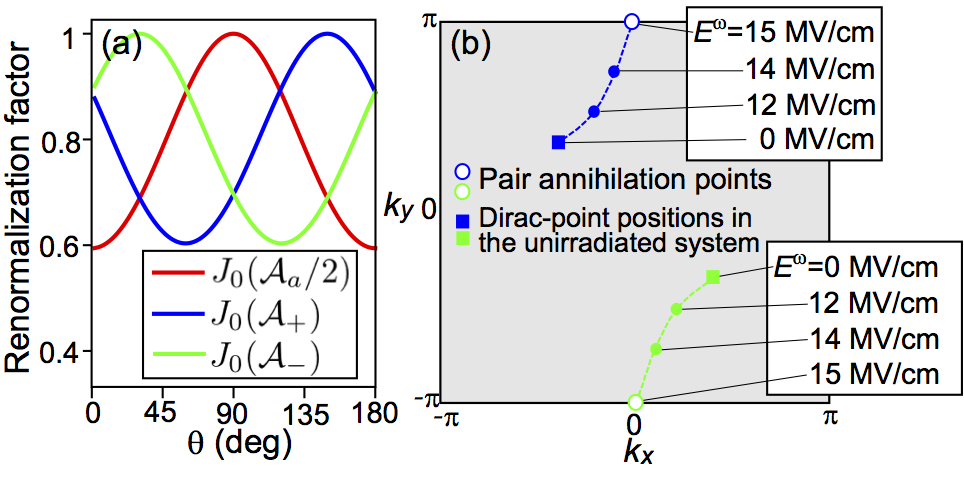}
\caption{(a) Photoinduced renormalization factors $J_0(\mathcal{A}_a/2)$, $J_0(\mathcal{A}_+)$ and $J_0(\mathcal{A}_-)$ for the transfer integrals in $\alpha$-(BEDT-TTF)$_2$I$_3$ irradiated with linearly polarized light as functions of the light polarization angle $\theta$. The light amplitude and frequency are fixed at $E^\omega$=15 MV/cm and $\hbar\omega$=0.6 eV, respectively. (b) Trajectories of the positive and negative emergent magnetic charges at the Dirac points with increasing light amplitude $E^\omega$ in the momentum space.}
\label{Fig5}
\end{figure}
%%%%%%%%%%%%%%%%%%%%%%%%%%%%%%%%
In the phase diagram of Fig.~\ref{Fig3}(a), the normal insulator phase appears only around $\theta$=45$^\circ$. This indicates that only linearly polarized light with a polarization angle of $\theta \sim 45^\circ$ gives rise to the pair annihilation of emergent magnetic charges.
% accompanied by the photoinduced phase transition from the Dirac semimetal phases ($E^\omega < 15$ MV/cm) to the normal insulator phase ($E^\omega > 15$ MV/cm) as the light amplitude $E^\omega$ increases.
This sensitivity to the polarization angle $\theta$ might be attributed to the anisotropic renormalizations of the transfer integrals. Under the irradiation of light, the transfer integrals are renormalized by factors represented by the Bessel functions (Eq.~(\ref{renorm})). More specifically, the transfer integrals for bonds along [010], [110] and [1$\bar{1}$0] directions are renormalized by factors $J_0(\mathcal{A}_a/2)$, $J_0(\mathcal{A}_+)$ and $J_0(\mathcal{A}_-)$, respectively (see Table.~\ref{tab:trenorm}). Because $\mathcal{A}_a$, $\mathcal{A}_+$ and $\mathcal{A}_-$ are functions of $\theta$ as seen in Eqs.~(\ref{eq:Aij2}) and (\ref{eq:Aij3}), the extent of the renormalization strongly depends on the angle $\theta$. In Fig.~\ref{Fig5}(a), we plot the three renormalization factors $J_0(\mathcal{A}_a/2)$, $J_0(\mathcal{A}_+)$ and $J_0(\mathcal{A}_-)$ as functions of $\theta$. We find that $J_0(\mathcal{A}_+)$ takes a minimum at $\theta \sim 45^\circ$, which means that the transfer integrals for bonds along the [110] direction are strongly suppressed at $\theta$=45$^\circ$. This anisotropic renormalization of the transfer integrals is expected to modulate the band dispersions along the [110] direction in the momentum space, which displaces the Dirac points along this direction. 
% and the resulting collision and pair annihilation of the emergent magnetic charges with opposite signs. 
To confirm this, we show the positions of the two emergent magnetic charges in Fig.~\ref{Fig5}(b). As the light amplitude $E^\omega$ increases from 0 to 15 MV/cm, the positions of the positive and negative magnetic charges move toward $(k_x, k_y)$=$(0, \pi)$ and $(0, -\pi)$, respectively. Notably, they move approximately along the [110] direction in the momentum space, which supports the idea that the pair annihilation results from the anisotropic renormalizations of transfer integrals. 

%%%%%%%%%%%%%%%%%%%%%%%%%%%%%%%
\begin{figure*} %[t\end{figure*}hb]
\includegraphics[scale=0.5]{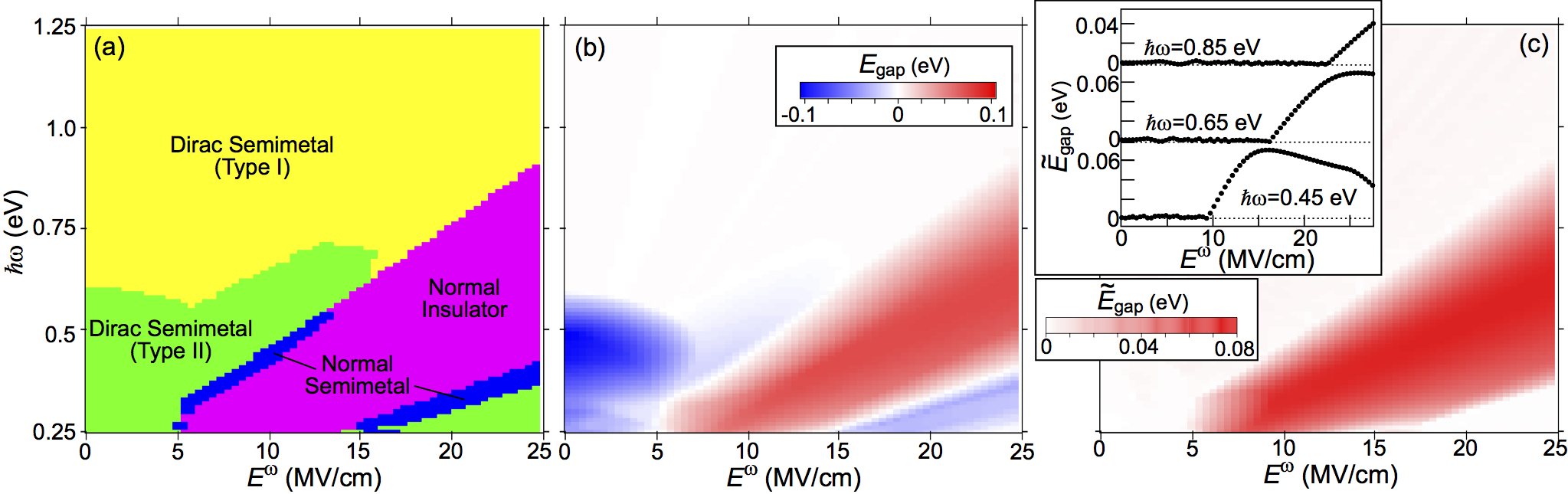}
\caption{(a) Phase diagram for nonequilibrium steady states in the photodriven $\alpha$-(BEDT-TTF)$_2$I$_3$ under irradiation with linearly polarized light in the plane of the amplitude $E^\omega$ and the frequency $\omega$ of light. (b)--(c) Color maps of the calculated two types of energy gap, (b) $E_{\rm gap}$ and (c) $\tilde{E}_{\rm gap}$ defined in Eqs.~(\ref{eq:Egap1}) and (\ref{eq:Egap2}), respectively. The inset of (c) shows the $E^\omega$ dependence of $\tilde{E}_{\rm gap}$ for selected light frequencies. The polarization angle $\theta$ of light is fixed at $\theta$=45$^\circ$.}
\label{Fig6}
\end{figure*}
%%%%%%%%%%%%%%%%%%%%%%%%%%%%%%%%
Figure~\ref{Fig6}(a) presents the phase diagram for nonequilibrium steady states in $\alpha$-(BEDT-TTF)$_2$I$_3$ irradiated with linearly polarized light in the plane of the amplitude $E^\omega$ and the frequency $\hbar\omega$ of light. When the light frequency is relatively low (high) as $\hbar\omega \lesssim 0.6$ eV ($\hbar\omega \gtrsim 0.75$ eV), the system enters the normal insulator phase or normal semimetal phase from the type-II (type-I) Dirac semimetal phase with increasing light amplitude $E^\omega$. In the intermediate frequency regime ($\hbar\omega\sim0.7$ eV), the system exhibits successive transitions from the type-I Dirac semimetal phase to the type-II Dirac semimetal phase to the normal insulator phase. The pair annihilation of the emergent magnetic charges occurs at the transition point from the Dirac semimetal phase to the normal insulator or the normal semimetal phase. We find that the light amplitude $E^\omega$ required to realize the pair annihilation increases almost linearly with the light frequency.

\section{Discussion}
\subsection{Band structure and off-resonant condition}
It is noted that in the present study, the photoinduced nonequilibrium steady phases were classified according to the features of the Floquet band structures, whereas the band occupations were not considered for the classification. In fact, when the light frequency $\hbar\omega$ is around 0.5--0.8 eV, a series of the bands ($\nu$=1,2,3,4) for the zero-photon states, which are located near the Fermi level, are separated from bands for the one-photon-absorbed ($n=+1$) states and those for the one-photon-emitted ($n=-1$) states (where this situation is referred to as the off-resonant situation), and the nonequilibrium band-occupation function $f_{n\nu}(\bm k)$ for the $\nu$th Floquet band approximately coincides with the Fermi--Dirac distribution function in the equilibrium as $f_{n\nu}(\bm{k}) \sim f_{\mathrm{FD}}(\varepsilon^n_{\nu}(\bm{k}))\delta_{n,0}$. In this case, our classification based on the Floquet band structure is justified.

Meanwhile, when the light frequency $\hbar\omega$ deviates from the range $0.5\lesssim\hbar\omega{\rm (eV)}\lesssim0.8$, the Floquet bands for different photon-number states overlap (this situation is referred to as the on-resonant situation). In such a situation, the band occupation deviates from the Fermi--Dirac distribution function. Therefore, even the normal insulator phase has nonzero conductivity in the on-resonant situation with lower light frequencies. The conductivity and transport properties in on-resonant systems should be calculated using the Floquet--Keldysh method~\cite{Tsuji09,Aoki14}, which is formulated by combining the Keldysh Green's function technique~\cite{Jauho94,Mahan00} and the Floquet theory. In contrast, when the system is in the off-resonant situation with $\hbar\omega$=0.5--0.8 eV, the normal insulator phase is indeed insulating.

The frequency window of the off-resonant condition (i.e., $0.5\lesssim\hbar\omega{\rm (eV)}\lesssim0.8$) for $\alpha$-(BEDT-TTF)$_2$I$_3$ is determined by the bandwidth $W$ of the band set ($\nu$=1,2,3,4) around the Fermi level relevant to electrons in the BEDT-TTF layer (i.e., the BEDT-TTF bands) and the energy spacing $G$ between the Fermi level and the upper/lower bands. The first-principles calculation in Ref.~\cite{Kino06} showed that both $W$ and $G$ are $\sim$0.8 eV in the static case. In the photodriven system, the bandwidth $W$ is renormalized as $\sim$0.5 eV for the typical light amplitudes and frequencies examined in the present study. Note that a set of the BEDT-TTF bands of the zero-photon states around the Fermi level overlaps that of the one-photon-absorbed (-emitted) states when $\hbar\omega\lesssim 0.5$ eV, whereas it overlaps the lower (upper) bands of the one-photon-absorbed (-emitted) states when $\hbar\omega\gtrsim 0.8$ eV, resulting in the on-resonant situation. The BEDT-TTF bands near the Fermi level are well separated from the upper and lower bands in this organic compound, which provides a rare opportunity to have a finite light-frequency window to realize the off-resonant situation. Thereby, $\alpha$-(BEDT-TTF)$_2$I$_3$ is a precious example material for studying photoinduced nonequilibrium phases and photoinduced phase transitions because of this peculiar band structure as well as the charge ordering as a source of the staggered site potential necessary for the predicted pair annihilation phenomenon. We expect that slight overlaps of the bands with different photon numbers (i.e., the weak on-resonant situation) never alter the band occupation so drastically from the equilibrium case that the phase classifications in Fig.~\ref{Fig6}(a), based on the Floquet band structure, are valid to some extent even above and below this frequency window.

\subsection{Experimental feasibility}
We now discuss the feasibility of experimentally observing the predicted photoinduced pair annihilation. Our quantitative predictions indicate that a rather strong light electric field of $E^\omega$$\sim$15 MV/cm is required to realize the pair annihilation. We know that samples may be damaged or even broken under continuous irradiation with an intense light field, but it is difficult to discuss to what extent samples of the organic compound can endure an intense light field. However, we consider that the experiment is worth trying or even feasible for the following reasons. First, several experiments of photoinduced phase transitions have been successfully performed for similar organic materials, such as $\kappa$-type BEDT-TTF compounds, at least with few-cycle-pulse or one-cycle-pulse laser light as intense as $E^\omega=$16 MV/cm~\cite{Kawakami18,Kawakami20}. Second, although a continuous-wave photoexcitation was assumed in the present study in applying the Floquet theorem, it has been experimentally demonstrated that continuous-wave photoirradiation is not necessarily required to observe the nonequilibrium steady states or the Floquet states, and a small-number-cycle pulse or even a less-than-10-cycle pulse is sufficient~\cite{Uchida15,Uchida16}. Third, it has been theoretically shown that because of electron correlation effects, the positions of the Dirac points in real materials are closer than those predicted using the present tight-binding model without electron correlations. Therefore, the pair annihilation may be realized with a light electric field weaker than the predicted field strength of $\sim$15 MV/cm~\cite{Kobayashi07,Ohki20}. Of course, there may be difficulties in conducting real experiments, but we expect that the predicted pair annihilation of the emergent magnetic charges will be observed experimentally in the near future as the difficulties are overcome.

\section{Conclusion}
We theoretically proposed a possible pair annihilation of emergent magnetic charges with opposite signs in the photodriven organic conductor $\alpha$-(BEDT-TTF)$_2$I$_3$ irradiated with linearly polarized light, which occurs in the momentum space as a consequence of the photoinduced phase transition from the Dirac semimetal phase to the normal insulator phase. This nonequilibrium phase transition is non-topological and caused by a gap opening at the Dirac points as a consequence of a band-structure deformation due to the photoinduced transfer-integral renormalizations. This non-topological photoinduced phase transition is distinct from the well-argued photoinduced topological phase transition to the Chern insulator phase due to a gap opening at the Dirac points under irradiation with circularly polarized light~\cite{Mikami16,Oka09,Kitagawa11,Lindner11,McIver19}. In previous work to date, emergent magnetic charges or magnetic (anti)monopoles originating from the topology of spin textures in a chiral magnet have been observed, and their pair annihilation has been argued to occur under the application of an external magnetic field~\cite{Kanazawa16}. Our work proposed, to our knowledge, the first example of a possible pair annihilation of emergent magnetic charges originating from the topology of the electronic band structures in the momentum space. We also discussed the feasibility of experimental observations of the predicted phases and phenomena. The photoinduced pair annihilation of magnetic charges and rich nonequilibrium steady phases in the photodriven $\alpha$-(BEDT-TTF)$_2$I$_3$ are expected to be observed in future experiments.

\section{Acknowledgments}
KK is supported by the World-leading Innovative Graduate Study Program for Materials Research, Industry, and Technology (MERIT-WINGS) of the University of Tokyo. MM is supported by JSPS KAKENHI (Grants No. 16H06345, No. 19H00864, No. 19K21858 and No. 20H00337), CREST, the Japan Science and Technology Agency (Grant No. JPMJCR20T1), a Research Grant in the Natural Sciences from the Mitsubishi Foundation, and a Waseda University Grant for Special Research Projects (Project No. 2020C-269). YT is supported by JSPS KAKENHI (Grants No. 19K23427 and No. 20K03841). MO is supported by JSPS KAKENHI (Grant No. 18H01162).

\end{document}